\documentclass[aps,prl,twocolumn,showpacs,superscriptaddress,floatfix]{revtex4-1}
\usepackage{amsmath,amssymb}
\usepackage{graphicx}
\usepackage{epsfig}
\usepackage{color}
\usepackage{rotating}
\usepackage{bm}
\usepackage{setspace}
\begin{document}
%Title of paper
\title{Charge density waves and Fermi level pinning in monolayer and bilayer SnSe$_2$}
\author{Shu-Ze Wang}
\author{Yi-Min Zhang}
\author{Jia-Qi Fan}
\author{Ming-Qiang Ren}
\affiliation{State Key Laboratory of Low-Dimensional Quantum Physics, Department of Physics, Tsinghua University, Beijing 100084, China}
\author{\\Can-Li Song}
\email[]{clsong07@mail.tsinghua.edu.cn}
\author{Xu-Cun Ma}
\email[]{xucunma@mail.tsinghua.edu.cn}
\affiliation{State Key Laboratory of Low-Dimensional Quantum Physics, Department of Physics, Tsinghua University, Beijing 100084, China}
\affiliation{Frontier Science Center for Quantum Information, Beijing 100084, China}
\author{Qi-Kun Xue}
\email[]{qkxue@mail.tsinghua.edu.cn}
\affiliation{State Key Laboratory of Low-Dimensional Quantum Physics, Department of Physics, Tsinghua University, Beijing 100084, China}
\affiliation{Frontier Science Center for Quantum Information, Beijing 100084, China}
\affiliation{Beijing Academy of Quantum Information Sciences, Beijing 100193, China}
\begin{abstract}
Materials with reduced dimensionality often exhibit exceptional properties that are different from their bulk counterparts. Here we report the emergence of a commensurate 2 $\times$ 2 charge density wave (CDW) in monolayer and bilayer SnSe$_2$ films by scanning tunneling microscope. The visualized spatial modulation of CDW phase becomes prominent near the Fermi level, which is pinned inside the semiconductor band gap of SnSe$_2$. We show that both CDW and Fermi level pinning are intimately correlated with band bending and virtual induced gap states at the semiconductor heterointerface. Through interface engineering, the electron-density-dependent phase diagram is established in SnSe$_2$. Fermi surface nesting between symmetry inequivalent electron pockets is revealed to drive the CDW formation and to provide an alternative CDW mechanism that might work in other compounds.
\end{abstract}

%\maketitle must follow title, authors, abstract, \pacs, and \keywords
\maketitle
\begin{spacing}{1.005}
Charge density wave (CDW) represents a static modulation of conduction electrons that is commonly accompanied by a periodic lattice distortion \cite{monceau2012electronic}. The understanding of such a collectively ordered electronic state has been widely supposed to be a key to decipher the secrets of unconventional superconductivity in many two-dimensional (2D) layered materials due to their close proximity in the electronic phase diagram \cite{keimer2015quantum, Isobe2018unconventional}. In quest of this objective, dimensionality and interface engineering have been often considered as unique strategies to create and control the CDW phase \cite{Calandra2009effect, soumyanarayanan2013quantum, lin2020patterns, frano2016long}. As a matter of fact, newly emerging or enhanced CDW correlations have been revealed in several transition metal dichalcogenides at the 2D limit \cite{Peng2015molecular, xi2015strongly, chen2017emergence, duvjir2018emergence}, whereas the interfacial effects on the CDW order of these 2D nanosheets were relatively little investigated \cite{Lefcochilos2019substrate, jia2018epitaxial}. A variety of scenarios invoking the saddle-point singularities \cite{Rice1975new}, Fermi surface nesting \cite{monceau2012electronic, duvjir2018emergence}, electron-phonon coupling \cite{xi2015strongly}, excitonic insulator \cite{wilson1977concerning, kogar2017signatures} and Jahn-Teller band instabilities \cite{chen2017emergence, whangbo1992analogies}, as well as some combination of them \cite{Kidd2002electron}, have been employed to account for the formation of CDW. However, a consensus on which factors play the primary roles in driving the CDW transition remains a hotly debated topic.

On the other hand, layered main-group metal dichalcogenides like SnSe$_2$ have recently attracted substantial interest owing to their potential applications in field-effect transistors \cite{guo2016field}, optoelectronic \cite{zhou2015ultrathin, huang2015designing} and thermoelectric devices \cite{luo2018n}, while the high abundance and low toxicity of Sn hold promises for the commercial use. Interface engineering via organometallic intercalation \cite{wu2019spacing, Song2019superconductivity}, heterostructure design \cite{Zhang2018observation, shao2019strongly} and dielectric gating techniques \cite{zeng2018gate} has led to the emergence of superconductivity in SnSe$_2$, although its bulk counterpart is well-known as semiconductor with an indirect band gap of 1.07 eV \cite{Gonzalez2016layer}. Moreover, periodic lattice distortions, possibly associated with CDW, were ever reported in pressurized and compressed SnSe$_2$ \cite{shao2019strongly, Ying2018unusual}. Astonishingly, the so-called CDW orders exhibit different wave vectors, calling for further study. Layered 1\textit{T}-SnSe$_2$ has thus become a newly-fertile playground for the exploration and manipulation of these many-body collective phenomena, as well as the interplay between them. In this study, we employ scanning tunneling microscopy (STM) to exploit interface-induced CDW order and Fermi level ($E_\textit{F}$) pinning in monolayer (ML) and bilayer (BL) SnSe$_2$ films prepared on Sn-terminated Si(111) and SrTiO$_3$(001) substrates. Virtual induced gap states (VIGSs) at the semiconductor heterointerface are visualized and found to correlate intimately with the occurrence of CDW and $E_\textrm{F}$ pinning in ultrathin SnSe$_2$.

All experiments were carried out in an ultrahigh vacuum cryogenic STM system (Unisoku), which is connected to a molecular beam epitaxy (MBE) for \textit{in-situ} sample preparation. The base pressure of both chambers is better than 1.0 $\times$ 10$^{-10}$ Torr. Arsenic-doped Si(111) wafers were cleaned by repeated flashing to 1200$^\circ$C, leading to a reconstructed Si(111)-$7\times$7 surface, while niobium-doped SrTiO$_3$(001) substrates were heated at 1200$^\circ$C for 20 min to get a clean surface. High purity Sn (99.9999$\%$) and Se (99.999$\%$) sources are co-evaporated from standard effusion cells onto the substrates at 210$^\circ$C \cite{Zhang2018observation}. Due to the very volatile nature of Se molecules from the effusion cell, a high Se/Sn flux ratio of $\sim$ 10 was used to compensate for the Se losses during the MBE growth, bearing a similar growth dynamic to that for other metal selenides \cite{song2011molecular}. After the film growth, the samples were immediately transferred into the STM stage for data collection at 4.5 K, unless otherwise specified. Polycrystalline PtIr tips were conditioned by electron beam heating, calibrated on Ag/Si(111) films and used throughout the experiment. Tunneling conductance spectra were measured by using a standard lock-in technique with a small bias modulation at 931 Hz.
\end{spacing}

Layered 1\textit{T}-SnSe$_2$ has a trigonal symmetry and consists of hexagonally packed Sn sheets sandwiched between the anionic Se sheets with an in-plane lattice parameter of approximately 3.81 $\textrm{\AA}$ \cite{Gonzalez2016layer}. In order to optimize the epitaxial growth of SnSe$_2$ thin films, freshly cut Si(111) substrates with an in-plane lattice parameter of 3.8403 $\textrm{\AA}$ are ideally chosen [Fig.\ 1(a)], yielding a lattice mismatch of $<0.8\%$. We further passivate the chemically reactive Si(111)-$7\times$7 substrates by depositing $\sim$ 1/3 ML Sn atoms at 600$^\circ$C, as detailed in the Supplemental Materials \cite{supplementary}. This results into a Sn-terminated Sn/Si(111)-$\sqrt{3}\times\sqrt{3}R30^\textrm{o}$ (hereafter referred to as $\sqrt{3}$-Sn/Si) surface, characteristic of a Mott-Hubbard insulating ground state [Figs.\ 1(b) and S1] \cite{Modesti2007insulating, Wu2020superconductivity}. Figure 1(c) depicts the three-dimensional (3D) Brillouin zone of SnSe$_2$ with high symmetry points and its 2D projection. Recent first-principles calculations and band structure measurements of SnSe$_2$ have located its conduction band minimum (CBM) and valence band maximum (VBM) along the $M$-$L$ and $\Gamma$-$M$($K$) directions \cite{Gonzalez2016layer,lochocki2019electronic}, respectively. This differs from the CDW-bearing sister compound 1\textit{T}-TiSe$_2$ with the VBM located justly at $\Gamma$ \cite{jia2018epitaxial}. Besides, TiSe$_2$ has been widely considered to be a semimetal or a narrow-gap semiconductor \cite{wilson1977concerning, Kidd2002electron, lochocki2019electronic}, whereas the semiconducting SnSe$_2$ exhibits a gap greater than 1.0 eV \cite{Gonzalez2016layer}. Despite these distinctions, ML and BL SnSe$_2$ films grown on $\sqrt{3}$-Sn/Si exhibit a clear charge order [Figs.\ 1(d) and 1(e)]. In analogy to 1\textit{T}-TiSe$_2$, the CDW modulation displays a commensurate 2 $\times$ 2 structure. More remarkably, the CDW vanishes in triple-layer (TL) SnSe$_2$, leaving behind an intact SnSe$_2$(001)-1 $\times$ 1 surface [Fig.\ 1(f)].

\begin{figure}[t]
\includegraphics[width=\columnwidth]{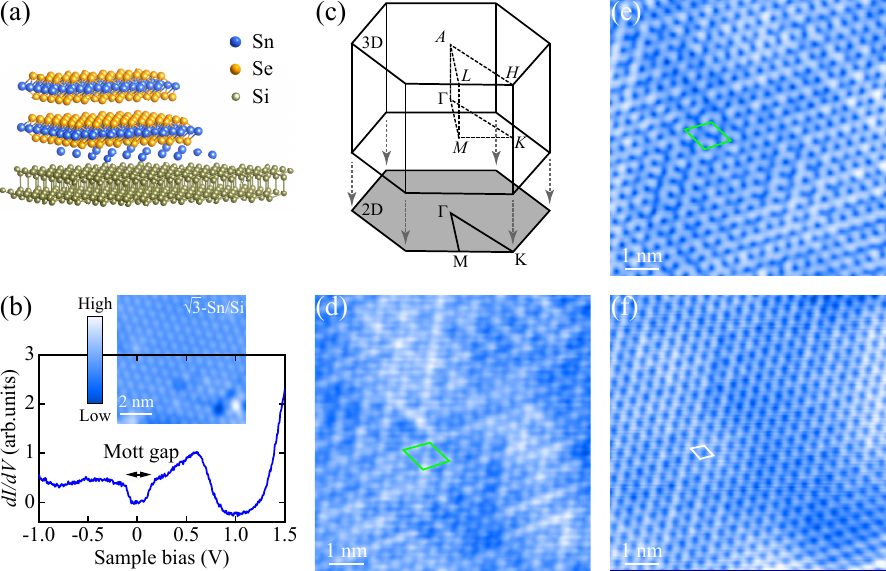}
\caption{(color online) (a) Sketch of 1\textit{T}-SnSe$_2$ (space group:\ P\={3}m1) films on Sn-terminated Si(111). (b) Differential conductance \textit{dI/dV} spectrum showing a Mott-Hubbard gap of $\sim$ 0.26 eV around $E_\textrm{F}$ in $\sqrt{3}$-Sn/Si. The set point is stabilized at $V$ = 1.6 V and $I$ = 100 pA. Inserted is a representative STM image of the $\sqrt{3}$-Sn/Si surface (8 nm $\times$ 8 nm, $V$ = $-$1.8 V, $I$ = 100 pA). (c) 3D and 2D Brillouin zones of SnSe$_2$. (d-f) STM topographies of ML, BL and TL SnSe$_2$ films epitaxially grown on $\sqrt{3}$-Sn/Si (8 nm $\times$ 8 nm), respectively. The white and green rhombuses mark the primitive and charge order unit cells, respectively. Imaging conditions are $V$ = 0.5 V and $I$ = 100 pA, except for (e) $V$ = 0.2 V.
}
\end{figure}

Provided that the lattice mismatch between SnSe$_2$ and the Si(111) substrate is negligibly small, we exclude that the observed CDW phase origins from any possible effects associated with the epitaxial strain \cite{shao2019strongly}. To understand the microscopic cause of the CDW order, we show a series of bias-dependent STM topographies and the corresponding fast Fourier transform (FFT) images in Figs.\ 2 and S2. Evidently, the CDW spots, marked by the green circles, become more prominent near $E_\textrm{F}$. This is understandable since a CDW phase transition mainly affects the density of states (DOS) of SnSe$_2$ close to $E_\textrm{F}$. More interestingly, the CDW wave vector $q_{\textrm{CDW}}$ reduces below $-$0.1 eV [Fig.\ S3], leading to a crossover from commensurate to incommensurate CDW state.

Figure 3(a) plots the CDW intensity (top panel), calculated as integrated FFT magnitude from the green circled regions in Fig.\ 2(b), as a function of the bias. As expected, the CDW is suppressed at an elevated temperature of 78 K. Compared to BL SnSe$_2$, ML SnSe$_2$ displays more pronounced CDW. This hints at the essential importance of interfacial effects in the CDW formation. A careful investigation of SnSe$_2$ thickness-dependent \textit{dI/dV} spectra in the bottom panel of Fig.\ 3(a) reveals an enhancement of the direct band gap E$_\textrm{g}^{\textrm{dir}}$($\Gamma$) upon reduction of the film thickness. This is consistent with a previous report \cite{Zhang2018observation} and could be accounted for in terms of the poor electrostatic screening and enhanced quantum confinement in few-layer SnSe$_2$ \cite {Gonzalez2016layer}. More significantly, we find that on the surface $E_\textrm{F}$ is pinned $\sim$ 0.4$-$0.5 eV below the CBM at $\Gamma$, irrespective of the SnSe$_2$ film thickness.

\begin{figure*}[t]
\includegraphics[width=2\columnwidth]{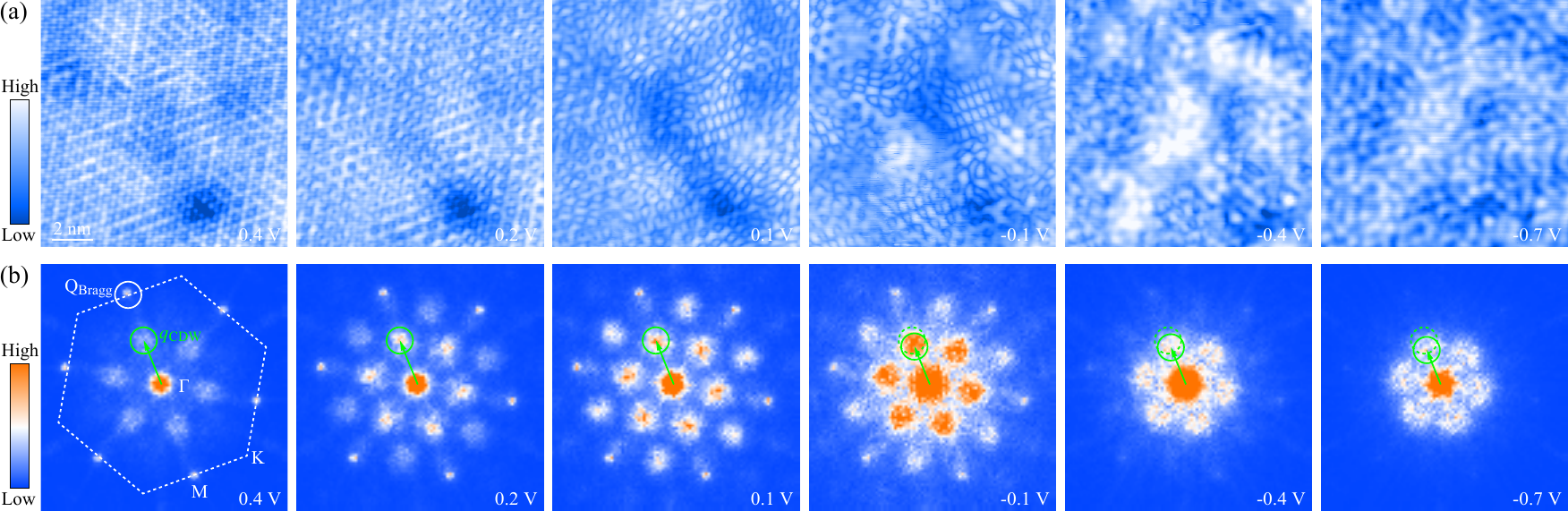}
\caption{(color online) (a) Constant-current STM topography images of monolayer SnSe$_2$/$\sqrt{3}$-Sn/Si films in the same field of view (12 nm $\times$ 12 nm, $I$ =100 pA) and (b) the corresponding FFT amplitudes at the indicated sample biases. Every FFT image has been sixfold symmetrized to optimize the signal-to-noise ratio. The Bragg peaks and CDW spots of SnSe$_2$ are white- and green-circled, respectively. The white hexagon in the left FFT image corresponds to the 2D Brillouin zone of SnSe$_2$, while the three dashed circles denote the 2 $\times$ 2 CDW spots just as the FFT images on the left. Note that the green arrows mark the CDW wave vectors that gradually decrease below $-$0.1 eV.
}
\end{figure*}

\begin{figure}[t]
\includegraphics[width=1\columnwidth]{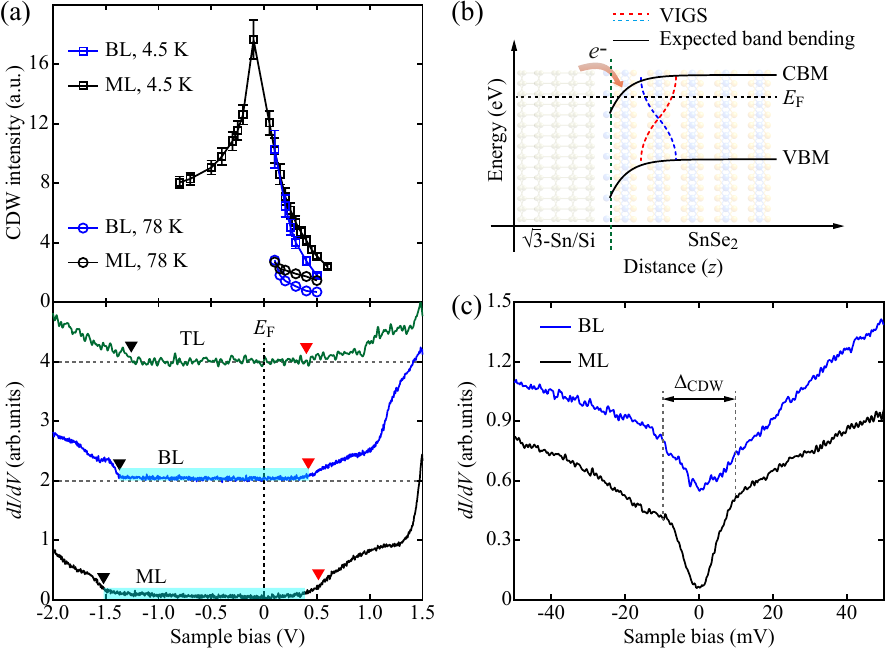}
\caption{(color online) (a) Energy-dependent CDW intensity (top panel) and wide-energy-scale \textit{dI/dV} spectra ($V$ = 1.5 V, $I$ = 100 pA) at varied film thicknesses (bottom panel). The black (red) triangles mark the VBM (CBM) near (at) $\Gamma$, with their spacing corresponding to the direct band gap E$_\textrm{g}^{\textrm{dir}}$($\Gamma$) of SnSe$_2$. The vertical dashed line denotes $E_\textrm{F}$ throughout, and the cyan rectangle indicates emergent in-gap states. (b) Energy band diagram for the SnSe$_2$/$\sqrt{3}$-Sn/Si semiconductor heterointerface, as well as the interfacial origin of the in-gap states. (c) Tunneling spectra showing CDW energy gaps near $E_\textrm{F}$ on both ML and BL SnSe$_2$.
}
\end{figure}

\begin{figure}[t]
\includegraphics[width=\columnwidth]{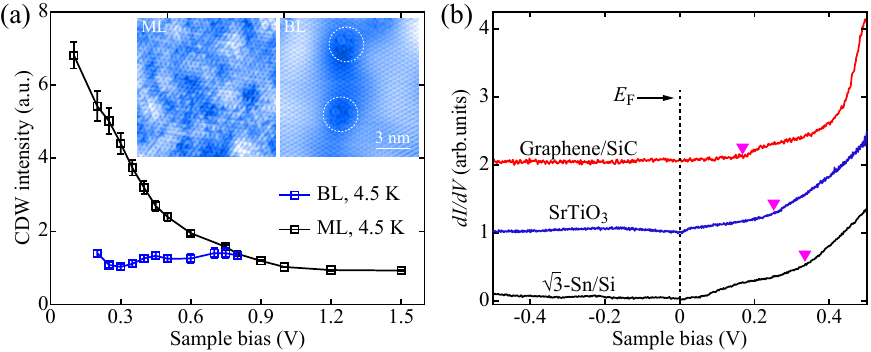}
\caption{(color online) (a) Calculated CDW intensity of ML and BL SnSe$_2$ films on SrTiO$_3$ substrate. Insets are STM images (12 nm $\times$ 12 nm, $V$ = 0.3 V, $I$ = 100 pA) of ML and BL SnSe$_2$ films on SrTiO$_3$, with the dashed circles embracing the short-range CDW in the vicinity of defects. (b) Comparison among \textit{dI/dV} spectra ($V$ = 0.5 V, $I$ = 100 pA) of ML SnSe$_2$ films on different substrates, measured in an energy range of $\pm$0.5 eV. The magenta triangles represent the CBM at the M point of 2D Brillouin zone in SnSe$_2$.
}
\end{figure}

The formation of CDW phase and $E_\textrm{F}$ pinning in ultrathin SnSe$_2$ films are closely related to emergent electronic states in the semiconducting gap, i.e.\ the cyan-marked finite DOS in Fig.\ 3(a). The in-gap states get suppressed with thickness and could not be ascribed to any impurity-induced bound states, because there exists little defect on the SnSe$_2$ films investigated [Figs.\ 1(d)-(f)]. Knowing that SnSe$_2$ has a large electron affinity \cite{zhang2018band}, upon contact electrons would flow from $\sqrt{3}$-Sn/Si to SnSe$_2$, leading to downward band bending of the SnSe$_2$ and confinement of 2D electron gas (2DEG) at the interface [Fig.\ 3(b)]. This might contribute to some in-gap states predominantly near the CBM at the M points of the 2D Brillouin zone. More significantly, VIGSs will develop at the SnSe$_2$/$\sqrt{3}$-Sn/Si semiconductor heterointerface and extend over the whole indirect band gap of SnSe$_2$ \cite{luth2001solid}. The VIGSs correspond to the imaginary components of the complex wave function and exponentially decay away from the interface \cite{luth2001solid}. As predicted, the decay lengths critically rely on the energy and diverge at the conduction and valance band edges for the VIGSs with conduction-band character and valance-band character, respectively. As a consequence, more conduction (valance)-derived VIGSs emerge around CBM and VBM, marked by the red (blue) dashed lines in Fig.\ 3(b). Based on the one-dimensional VIGS model \cite {monch1990physics}, the minimum decay length of VIGSs occurs roughly at the midgap energy and approximates to 6.0 $\textrm{\AA}$ for SnSe$_2$. This is consistent with the sharply suppressed in-gap states on BL and TL SnSe$_2$ (see the bottom panel of Fig.\ 3(a)). Plotted in Fig. 3(c) are the smaller-energy-scale tunneling spectra of ML and BL SnSe$_2$ films. Partial gaps associated with the CDW phase are identified in the vicinity of $E_\textrm{F}$. The gap magnitude $\Delta_{\textrm{CDW}}\sim$ 20 meV agrees quantitatively with the significant suppression of CDW at 78 K [Fig.\ 3(a)].

In order to provide more insights into the CDW phase of SnSe$_2$, we also prepare ultrathin SnSe$_2$ films on SrTiO$_3$ substrate. Distinct from a recent report \cite{shao2019strongly}, no apparent lattice compression is revealed. This seems more understandable because the MBE growth of layered SnSe$_2$ films is of quasi-van der Waals epitaxy. Shown in Fig.\ 4(a) are STM topographies and energy dependent CDW intensity in ML and BL SnSe$_2$/SrTiO$_3$ films. Evidently, the CDW modulations are weaker than those in SnSe$_2$/$\sqrt{3}$-Sn/Si. In particular, only short-ranged CDW order, circled by the white dashes in Fig.\ 4(a), could be seen to surround single native defects in BL SnSe$_2$/SrTiO$_3$ films.

\begin{figure}[t]
\includegraphics[width=0.9\columnwidth]{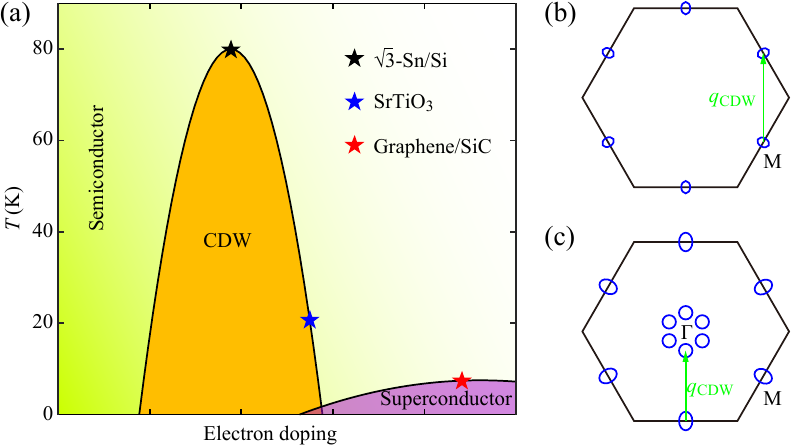}
\caption{(color online) (a) Electronic phase diagram of SnSe$_2$. The stars denote monolayer SnSe$_2$ films on $\sqrt{3}$-Sn/Si, SrTiO$_3$  and graphene/SiC substrates, respectively. (b) Fermi surface and (c) constant energy contour (i.e.\ the midgap energy level) of SnSe$_2$ films. The black hexagons denote the 2D Brillouin zone of SnSe$_2$, while green arrows the $q_{\textrm{CDW}}$ of CDW order.
}
\end{figure}

Figure 4(b) compares the tunneling \textit{dI/dV} spectra of monolayer SnSe$_2$ films on the indicated substrates, with the MBE growth of SnSe$_2$ on graphene/SiC substrate described in detail elsewhere \cite{Zhang2018observation}. The conductance kinks, marked by magenta triangles, correspond to the CBM at M. The emergence of VIGSs proves generic for all three heterostructures and serve as a reservoir for electrons/holes. These states effectively pin $E_\textrm{F}$ and hold the key for the CDW formation in ML and BL SnSe$_2$. Given that the band bending and VIGS decay length are determined primarily by the semiconductor parameters of SnSe$_2$ \cite {monch1990physics, kerelsky2017absence}, the smaller energy spacing between the CBM at the M points and $E_\textrm{F}$ means more electron doping. Such distinction may be probably related to the different work function of the substrates. Anyhow, our comparison study enables to establish a doping-dependent phase diagram of SnSe$_2$ in Fig.\ 5(a). Upon electron doping via interface engineering, the CDW develops with the filled VIGSs in ML and BL SnSe$_2$/$\sqrt{3}$-Sn/Si films. As the electron doping is further increased, the CDW phase gets suppressed in ultrathin SnSe$_2$/SrTiO$_3$ films and eventually evolves into a superconducting state when the SnSe$_2$ films are grown on a graphitized SiC(0001) substrate \cite{Zhang2018observation}. This complete electronic phase diagram highlights the importance of dimensionality and electron doping in SnSe$_2$ semiconductor, and offers fresh insights into the relationship between the CDW phase and superconductivity.

We now discuss possible mechanisms for the observed CDW phase. The large semiconductor band gap of 1.07 eV prevents an excitonic insulator scenario from being responsible for the CDW order in SnSe$_2$, where an exciton binding energy larger than the band gap is needed but hardly fulfilled \cite{wilson1977concerning}. This holds even more true for ultrathin SnSe$_2$ films with enhanced band gap [Fig.\ 3(a)]. Secondly, a Jahn-Teller band instability, as widely explored for explaining CDW in Ti-based dichalcogenides \cite{chen2017emergence, Rossnagel2002charge, rossnagel2010suppression}, could be safely excluded as well. This is primarily because a starting point of the band-type Jahn-Teller interaction is the substantial overlap of conduction and valance bands, which is yet separated by the semiconductor band gap in SnSe$_2$. Moreover, this model relates intimately to a low-lying d-orbital splitting \cite{hughes1977structural} or a s-d orbital mixing \cite{whangbo1992analogies}. Distinct from TiSe$_2$, however, the d orbital is little involved in the conduction and valance bands of the s-p metal dichalcogenide SnSe$_2$ \cite{Gonzalez2016layer}. As thus, our results call for other scenarios for the formation of CDW in SnSe$_2$. Obviously, any models associated with saddle-point singularities are unlikely, since $E_\textrm{F}$ of the SnSe$_2$ films is well pinned inside the band gap and no sharp DOS enhancement is observable around $E_\textrm{F}$ [Figs.\ 3(a) and 4(b)].

Next we resort to a classical concept of CDW order induced by Fermi surface nesting \cite{monceau2012electronic}. This is in agreement with our observation of CDW only in the presence of sufficient in-gap states or electronic DOS around $E_\textrm{F}$. Multilayered and bulk SnSe$_2$ without low-lying DOS exhibit no CDW. Since $E_\textrm{F}$ of ultrathin SnSe$_2$ films lies closer to the CBM(M) than VBM($\Gamma$), the electronic DOS in the vicinity of $E_\textrm{F}$ originates predominantly from the band bending confined 2DES and VIGSs with conduction-band character at M, illustrated in Fig.\ 5(b). There exists little DOS at $\Gamma$, and thus a simple Fermi surface nesting connecting the CBM(M) than VBM($\Gamma$) seems unlikely. Alternatively, the Fermi surface nesting between symmetry inequivalent electron pockets at M, marked by the green arrow in Fig.\ 5(b), might be the driving force for the CDW formation. This is especially true as the electron pockets are small and the Fermi surface can nest nicely at the wave vector of the commensurate 2 $\times$ 2 CDW order observed. Here the elastic energy cost of lattice modulation is compensated by the total electronic energy gain via opening a gap at $E_\textrm{F}$ [Fig.\ 3(c)] and pushing the nearby states to lower energies. Upon additional electron doping, the electron pockets become larger, deteriorating the Fermi surface nesting for the CDW phase. As the CDW phase is almost killed due to the poor Fermi surface nesting, the superconductivity occurs probably associated with the intrapocket scattering at M. They seem fitting neatly to the electronic phase diagram in Fig.\ 5(a).

Finally, we note that as the applied bias is lowered far below $E_\textrm{F}$, i.e.\ roughly at the midgap energy of $-$0.7 eV [Fig.\ 2], VIGSs with the valance-band character develop equivalently near $\Gamma$, prompting substantial scattering between the $\Gamma$- and M-near VIGSs [Fig.\ 5(c)] at the given energy. Provided that the VBM and thus the valance-band-derived DOS are not strictly located at $\Gamma$ \cite{Gonzalez2016layer,lochocki2019electronic}, the scattering wave vector $q_{\textrm{CDW}}$ should be smaller than that of 2 $\times$ 2 order. This will induce a commensurate-to-incommensurate transition of the CDW order, as observed above. The incommensurate CDW is found to run along the $\Gamma$-M direction [Fig.\ 2]. This indicates that the VBM orient along the $\Gamma$-M direction as well \cite{Gonzalez2016layer}, rather than the $\Gamma$-K direction.

Our detailed STM study has revealed a commensurate 2 $\times$ 2 CDW phase in ultrathin SnSe$_2$ epitaxial films prepared on both $\sqrt{3}$-Sn/Si and SrTiO$_3$ substrates. Such a collectively ordered state is found to correlate intimately with the band bending and emergent VIGSs at the semiconductor heterointerfaces. We propose a novel mechanism associated with Fermi surface nesting between the symmetry inequivalent electron pockets at M to account for the formation of CDW nicely. The present method of interface engineering opens up possibilities in searching for novel states of matter at the 2D limit.

\begin{acknowledgments}
This work is financially supported by the National Natural Science Foundation of China (Grants No.\ 11774192, No.\ 11634007), the Ministry of Science and Technology of China (Grants No.\ 2016YFA0301004, No.\ 2017YFA0304600, No.\ 2018YFA0305603), and in part by the Beijing Advanced Innovation Center for Future Chip.
\end{acknowledgments}

% Create the reference section using BibTeX:
%\bibliography{SnSe2_CDW}

\begin{thebibliography}{41}%
\makeatletter
\providecommand \@ifxundefined [1]{%
 \@ifx{#1\undefined}
}%
\providecommand \@ifnum [1]{%
 \ifnum #1\expandafter \@firstoftwo
 \else \expandafter \@secondoftwo
 \fi
}%
\providecommand \@ifx [1]{%
 \ifx #1\expandafter \@firstoftwo
 \else \expandafter \@secondoftwo
 \fi
}%
\providecommand \natexlab [1]{#1}%
\providecommand \enquote  [1]{``#1''}%
\providecommand \bibnamefont  [1]{#1}%
\providecommand \bibfnamefont [1]{#1}%
\providecommand \citenamefont [1]{#1}%
\providecommand \href@noop [0]{\@secondoftwo}%
\providecommand \href [0]{\begingroup \@sanitize@url \@href}%
\providecommand \@href[1]{\@@startlink{#1}\@@href}%
\providecommand \@@href[1]{\endgroup#1\@@endlink}%
\providecommand \@sanitize@url [0]{\catcode `\\12\catcode `\$12\catcode
  `\&12\catcode `\#12\catcode `\^12\catcode `\_12\catcode `\%12\relax}%
\providecommand \@@startlink[1]{}%
\providecommand \@@endlink[0]{}%
\providecommand \url  [0]{\begingroup\@sanitize@url \@url }%
\providecommand \@url [1]{\endgroup\@href {#1}{\urlprefix }}%
\providecommand \urlprefix  [0]{URL }%
\providecommand \Eprint [0]{\href }%
\providecommand \doibase [0]{http://dx.doi.org/}%
\providecommand \selectlanguage [0]{\@gobble}%
\providecommand \bibinfo  [0]{\@secondoftwo}%
\providecommand \bibfield  [0]{\@secondoftwo}%
\providecommand \translation [1]{[#1]}%
\providecommand \BibitemOpen [0]{}%
\providecommand \bibitemStop [0]{}%
\providecommand \bibitemNoStop [0]{.\EOS\space}%
\providecommand \EOS [0]{\spacefactor3000\relax}%
\providecommand \BibitemShut  [1]{\csname bibitem#1\endcsname}%
\let\auto@bib@innerbib\@empty
%</preamble>
\bibitem [{\citenamefont {Monceau}(2012)}]{monceau2012electronic}%
  \BibitemOpen
  \bibfield  {author} {\bibinfo {author} {\bibfnamefont {P.}~\bibnamefont
  {Monceau}},\ }\href {\doibase 10.1080/00018732.2012.719674} {\bibfield
  {journal} {\bibinfo  {journal} {Adv. Phys.}\ }\textbf {\bibinfo {volume}
  {61}},\ \bibinfo {pages} {325} (\bibinfo {year} {2012})}\BibitemShut
  {NoStop}%
\bibitem [{\citenamefont {Keimer}\ \emph {et~al.}(2015)\citenamefont {Keimer},
  \citenamefont {Kivelson}, \citenamefont {Norman}, \citenamefont {Uchida},\
  and\ \citenamefont {Zaanen}}]{keimer2015quantum}%
  \BibitemOpen
  \bibfield  {author} {\bibinfo {author} {\bibfnamefont {B.}~\bibnamefont
  {Keimer}}, \bibinfo {author} {\bibfnamefont {S.}~\bibnamefont {Kivelson}},
  \bibinfo {author} {\bibfnamefont {M.}~\bibnamefont {Norman}}, \bibinfo
  {author} {\bibfnamefont {S.}~\bibnamefont {Uchida}}, \ and\ \bibinfo {author}
  {\bibfnamefont {J.}~\bibnamefont {Zaanen}},\ }\href {\doibase
  10.1038/nature14165} {\bibfield  {journal} {\bibinfo  {journal} {Nature}\
  }\textbf {\bibinfo {volume} {518}},\ \bibinfo {pages} {179} (\bibinfo {year}
  {2015})}\BibitemShut {NoStop}%
\bibitem [{\citenamefont {Isobe}\ \emph {et~al.}(2018)\citenamefont {Isobe},
  \citenamefont {Yuan},\ and\ \citenamefont {Fu}}]{Isobe2018unconventional}%
  \BibitemOpen
  \bibfield  {author} {\bibinfo {author} {\bibfnamefont {H.}~\bibnamefont
  {Isobe}}, \bibinfo {author} {\bibfnamefont {N.~F.~Q.}\ \bibnamefont {Yuan}},
  \ and\ \bibinfo {author} {\bibfnamefont {L.}~\bibnamefont {Fu}},\ }\href
  {\doibase 10.1103/PhysRevX.8.041041} {\bibfield  {journal} {\bibinfo
  {journal} {Phys. Rev. X}\ }\textbf {\bibinfo {volume} {8}},\ \bibinfo {pages}
  {041041} (\bibinfo {year} {2018})}\BibitemShut {NoStop}%
\bibitem [{\citenamefont {Calandra}\ \emph {et~al.}(2009)\citenamefont
  {Calandra}, \citenamefont {Mazin},\ and\ \citenamefont
  {Mauri}}]{Calandra2009effect}%
  \BibitemOpen
  \bibfield  {author} {\bibinfo {author} {\bibfnamefont {M.}~\bibnamefont
  {Calandra}}, \bibinfo {author} {\bibfnamefont {I.~I.}\ \bibnamefont {Mazin}},
  \ and\ \bibinfo {author} {\bibfnamefont {F.}~\bibnamefont {Mauri}},\ }\href
  {\doibase 10.1103/PhysRevB.80.241108} {\bibfield  {journal} {\bibinfo
  {journal} {Phys. Rev. B}\ }\textbf {\bibinfo {volume} {80}},\ \bibinfo
  {pages} {241108} (\bibinfo {year} {2009})}\BibitemShut {NoStop}%
\bibitem [{\citenamefont {Soumyanarayanan}\ \emph {et~al.}(2013)\citenamefont
  {Soumyanarayanan}, \citenamefont {Yee}, \citenamefont {He}, \citenamefont
  {Van~Wezel}, \citenamefont {Rahn}, \citenamefont {Rossnagel}, \citenamefont
  {Hudson}, \citenamefont {Norman},\ and\ \citenamefont
  {Hoffman}}]{soumyanarayanan2013quantum}%
  \BibitemOpen
  \bibfield  {author} {\bibinfo {author} {\bibfnamefont {A.}~\bibnamefont
  {Soumyanarayanan}}, \bibinfo {author} {\bibfnamefont {M.~M.}\ \bibnamefont
  {Yee}}, \bibinfo {author} {\bibfnamefont {Y.}~\bibnamefont {He}}, \bibinfo
  {author} {\bibfnamefont {J.}~\bibnamefont {Van~Wezel}}, \bibinfo {author}
  {\bibfnamefont {D.~J.}\ \bibnamefont {Rahn}}, \bibinfo {author}
  {\bibfnamefont {K.}~\bibnamefont {Rossnagel}}, \bibinfo {author}
  {\bibfnamefont {E.~W.}\ \bibnamefont {Hudson}}, \bibinfo {author}
  {\bibfnamefont {M.~R.}\ \bibnamefont {Norman}}, \ and\ \bibinfo {author}
  {\bibfnamefont {J.~E.}\ \bibnamefont {Hoffman}},\ }\href {\doibase
  10.1073/pnas.1211387110} {\bibfield  {journal} {\bibinfo  {journal} {Proc.
  Natl. Acad. Sci. USA}\ }\textbf {\bibinfo {volume} {110}},\ \bibinfo {pages}
  {1623} (\bibinfo {year} {2013})}\BibitemShut {NoStop}%
\bibitem [{\citenamefont {Lin}\ \emph {et~al.}(2020)\citenamefont {Lin},
  \citenamefont {Li}, \citenamefont {Wen}, \citenamefont {Berger},
  \citenamefont {Forr{\'o}}, \citenamefont {Zhou}, \citenamefont {Jia},
  \citenamefont {Taniguchi}, \citenamefont {Watanabe}, \citenamefont {Xi},\
  and\ \citenamefont {Bahramy}}]{lin2020patterns}%
  \BibitemOpen
  \bibfield  {author} {\bibinfo {author} {\bibfnamefont {D.~J.}\ \bibnamefont
  {Lin}}, \bibinfo {author} {\bibfnamefont {S.~C.}\ \bibnamefont {Li}},
  \bibinfo {author} {\bibfnamefont {J.~S.}\ \bibnamefont {Wen}}, \bibinfo
  {author} {\bibfnamefont {H.}~\bibnamefont {Berger}}, \bibinfo {author}
  {\bibfnamefont {L.}~\bibnamefont {Forr{\'o}}}, \bibinfo {author}
  {\bibfnamefont {H.~B.}\ \bibnamefont {Zhou}}, \bibinfo {author}
  {\bibfnamefont {S.}~\bibnamefont {Jia}}, \bibinfo {author} {\bibfnamefont
  {T.}~\bibnamefont {Taniguchi}}, \bibinfo {author} {\bibfnamefont
  {K.}~\bibnamefont {Watanabe}}, \bibinfo {author} {\bibfnamefont {X.~X.}\
  \bibnamefont {Xi}}, \ and\ \bibinfo {author} {\bibfnamefont {M.~S.}\
  \bibnamefont {Bahramy}},\ }\href {\doibase 10.1038/s41467-020-15715-w}
  {\bibfield  {journal} {\bibinfo  {journal} {Nat. Commun.}\ }\textbf {\bibinfo
  {volume} {11}},\ \bibinfo {pages} {1} (\bibinfo {year} {2020})}\BibitemShut
  {NoStop}%
\bibitem [{\citenamefont {Frano}\ \emph {et~al.}(2016)\citenamefont {Frano},
  \citenamefont {Blanco-Canosa}, \citenamefont {Schierle}, \citenamefont {Lu},
  \citenamefont {Wu}, \citenamefont {Bluschke}, \citenamefont {Minola},
  \citenamefont {Christiani}, \citenamefont {Habermeier}, \citenamefont
  {Logvenov}, \citenamefont {Wang}, \citenamefont {van Aken}, \citenamefont
  {Benckiser}, \citenamefont {Weschke}, \citenamefont {Tacon},\ and\
  \citenamefont {Keimer}}]{frano2016long}%
  \BibitemOpen
  \bibfield  {author} {\bibinfo {author} {\bibfnamefont {A.}~\bibnamefont
  {Frano}}, \bibinfo {author} {\bibfnamefont {S.}~\bibnamefont
  {Blanco-Canosa}}, \bibinfo {author} {\bibfnamefont {E.}~\bibnamefont
  {Schierle}}, \bibinfo {author} {\bibfnamefont {Y.}~\bibnamefont {Lu}},
  \bibinfo {author} {\bibfnamefont {M.}~\bibnamefont {Wu}}, \bibinfo {author}
  {\bibfnamefont {M.}~\bibnamefont {Bluschke}}, \bibinfo {author}
  {\bibfnamefont {M.}~\bibnamefont {Minola}}, \bibinfo {author} {\bibfnamefont
  {G.}~\bibnamefont {Christiani}}, \bibinfo {author} {\bibfnamefont {H.~U.}\
  \bibnamefont {Habermeier}}, \bibinfo {author} {\bibfnamefont
  {G.}~\bibnamefont {Logvenov}}, \bibinfo {author} {\bibfnamefont
  {Y.}~\bibnamefont {Wang}}, \bibinfo {author} {\bibfnamefont {P.~A.}\
  \bibnamefont {van Aken}}, \bibinfo {author} {\bibfnamefont {E.}~\bibnamefont
  {Benckiser}}, \bibinfo {author} {\bibfnamefont {E.}~\bibnamefont {Weschke}},
  \bibinfo {author} {\bibfnamefont {M.~L.}\ \bibnamefont {Tacon}}, \ and\
  \bibinfo {author} {\bibfnamefont {B.}~\bibnamefont {Keimer}},\ }\href
  {\doibase 10.1038/nmat4682} {\bibfield  {journal} {\bibinfo  {journal} {Nat.
  Mater.}\ }\textbf {\bibinfo {volume} {15}},\ \bibinfo {pages} {831} (\bibinfo
  {year} {2016})}\BibitemShut {NoStop}%
\bibitem [{\citenamefont {Peng}\ \emph {et~al.}(2015)\citenamefont {Peng},
  \citenamefont {Guan}, \citenamefont {Zhang}, \citenamefont {Song},
  \citenamefont {Wang}, \citenamefont {He}, \citenamefont {Xue},\ and\
  \citenamefont {Ma}}]{Peng2015molecular}%
  \BibitemOpen
  \bibfield  {author} {\bibinfo {author} {\bibfnamefont {J.~P.}\ \bibnamefont
  {Peng}}, \bibinfo {author} {\bibfnamefont {J.~Q.}\ \bibnamefont {Guan}},
  \bibinfo {author} {\bibfnamefont {H.~M.}\ \bibnamefont {Zhang}}, \bibinfo
  {author} {\bibfnamefont {C.~L.}\ \bibnamefont {Song}}, \bibinfo {author}
  {\bibfnamefont {L.}~\bibnamefont {Wang}}, \bibinfo {author} {\bibfnamefont
  {K.}~\bibnamefont {He}}, \bibinfo {author} {\bibfnamefont {Q.~K.}\
  \bibnamefont {Xue}}, \ and\ \bibinfo {author} {\bibfnamefont {X.~C.}\
  \bibnamefont {Ma}},\ }\href {\doibase 10.1103/PhysRevB.91.121113} {\bibfield
  {journal} {\bibinfo  {journal} {Phys. Rev. B}\ }\textbf {\bibinfo {volume}
  {91}},\ \bibinfo {pages} {121113} (\bibinfo {year} {2015})}\BibitemShut
  {NoStop}%
\bibitem [{\citenamefont {Xi}\ \emph {et~al.}(2015)\citenamefont {Xi},
  \citenamefont {Zhao}, \citenamefont {Wang}, \citenamefont {Berger},
  \citenamefont {Forr{\'o}}, \citenamefont {Shan},\ and\ \citenamefont
  {Mak}}]{xi2015strongly}%
  \BibitemOpen
  \bibfield  {author} {\bibinfo {author} {\bibfnamefont {X.~X.}\ \bibnamefont
  {Xi}}, \bibinfo {author} {\bibfnamefont {L.}~\bibnamefont {Zhao}}, \bibinfo
  {author} {\bibfnamefont {Z.~F.}\ \bibnamefont {Wang}}, \bibinfo {author}
  {\bibfnamefont {H.}~\bibnamefont {Berger}}, \bibinfo {author} {\bibfnamefont
  {L.}~\bibnamefont {Forr{\'o}}}, \bibinfo {author} {\bibfnamefont
  {J.}~\bibnamefont {Shan}}, \ and\ \bibinfo {author} {\bibfnamefont {K.~F.}\
  \bibnamefont {Mak}},\ }\href {\doibase 10.1038/nnano.2015.143} {\bibfield
  {journal} {\bibinfo  {journal} {Nat. Nanotechnol.}\ }\textbf {\bibinfo
  {volume} {10}},\ \bibinfo {pages} {765} (\bibinfo {year} {2015})}\BibitemShut
  {NoStop}%
\bibitem [{\citenamefont {Chen}\ \emph {et~al.}(2017)\citenamefont {Chen},
  \citenamefont {Pai}, \citenamefont {Chan}, \citenamefont {Takayama},
  \citenamefont {Xu}, \citenamefont {Karn}, \citenamefont {Hasegawa},
  \citenamefont {Chou}, \citenamefont {Mo}, \citenamefont {Fedorov},\ and\
  \citenamefont {Chiang}}]{chen2017emergence}%
  \BibitemOpen
  \bibfield  {author} {\bibinfo {author} {\bibfnamefont {P.}~\bibnamefont
  {Chen}}, \bibinfo {author} {\bibfnamefont {W.~W.}\ \bibnamefont {Pai}},
  \bibinfo {author} {\bibfnamefont {Y.~H.}\ \bibnamefont {Chan}}, \bibinfo
  {author} {\bibfnamefont {A.}~\bibnamefont {Takayama}}, \bibinfo {author}
  {\bibfnamefont {C.~Z.}\ \bibnamefont {Xu}}, \bibinfo {author} {\bibfnamefont
  {A.}~\bibnamefont {Karn}}, \bibinfo {author} {\bibfnamefont {S.}~\bibnamefont
  {Hasegawa}}, \bibinfo {author} {\bibfnamefont {M.~Y.}\ \bibnamefont {Chou}},
  \bibinfo {author} {\bibfnamefont {S.~K.}\ \bibnamefont {Mo}}, \bibinfo
  {author} {\bibfnamefont {A.~V.}\ \bibnamefont {Fedorov}}, \ and\ \bibinfo
  {author} {\bibfnamefont {T.~C.}\ \bibnamefont {Chiang}},\ }\href {\doibase
  10.1038/s41467-017-00641-1} {\bibfield  {journal} {\bibinfo  {journal} {Nat.
  Commun.}\ }\textbf {\bibinfo {volume} {8}},\ \bibinfo {pages} {1} (\bibinfo
  {year} {2017})}\BibitemShut {NoStop}%
\bibitem [{\citenamefont {Duvjir}\ \emph {et~al.}(2018)\citenamefont {Duvjir},
  \citenamefont {Choi}, \citenamefont {Jang}, \citenamefont {Ulstrup},
  \citenamefont {Kang}, \citenamefont {Thi~Ly}, \citenamefont {Kim},
  \citenamefont {Choi}, \citenamefont {Jozwiak}, \citenamefont {Bostwick},
  \citenamefont {Rotenberg}, \citenamefont {Park}, \citenamefont {Sankar},
  \citenamefont {Kim}, \citenamefont {Kim},\ and\ \citenamefont
  {Chang}}]{duvjir2018emergence}%
  \BibitemOpen
  \bibfield  {author} {\bibinfo {author} {\bibfnamefont {G.}~\bibnamefont
  {Duvjir}}, \bibinfo {author} {\bibfnamefont {B.~K.}\ \bibnamefont {Choi}},
  \bibinfo {author} {\bibfnamefont {I.}~\bibnamefont {Jang}}, \bibinfo {author}
  {\bibfnamefont {S.}~\bibnamefont {Ulstrup}}, \bibinfo {author} {\bibfnamefont
  {S.}~\bibnamefont {Kang}}, \bibinfo {author} {\bibfnamefont {T.}~\bibnamefont
  {Thi~Ly}}, \bibinfo {author} {\bibfnamefont {S.}~\bibnamefont {Kim}},
  \bibinfo {author} {\bibfnamefont {Y.~H.}\ \bibnamefont {Choi}}, \bibinfo
  {author} {\bibfnamefont {C.}~\bibnamefont {Jozwiak}}, \bibinfo {author}
  {\bibfnamefont {A.}~\bibnamefont {Bostwick}}, \bibinfo {author}
  {\bibfnamefont {E.}~\bibnamefont {Rotenberg}}, \bibinfo {author}
  {\bibfnamefont {J.~G.}\ \bibnamefont {Park}}, \bibinfo {author}
  {\bibfnamefont {R.}~\bibnamefont {Sankar}}, \bibinfo {author} {\bibfnamefont
  {K.~S.}\ \bibnamefont {Kim}}, \bibinfo {author} {\bibfnamefont
  {J.}~\bibnamefont {Kim}}, \ and\ \bibinfo {author} {\bibfnamefont {Y.~J.}\
  \bibnamefont {Chang}},\ }\href {\doibase 10.1021/acs.nanolett.8b01764}
  {\bibfield  {journal} {\bibinfo  {journal} {Nano Lett.}\ }\textbf {\bibinfo
  {volume} {18}},\ \bibinfo {pages} {5432} (\bibinfo {year}
  {2018})}\BibitemShut {NoStop}%
\bibitem [{\citenamefont {Lefcochilos-Fogelquist}\ \emph
  {et~al.}(2019)\citenamefont {Lefcochilos-Fogelquist}, \citenamefont
  {Albertini},\ and\ \citenamefont {Liu}}]{Lefcochilos2019substrate}%
  \BibitemOpen
  \bibfield  {author} {\bibinfo {author} {\bibfnamefont {H.~M.}\ \bibnamefont
  {Lefcochilos-Fogelquist}}, \bibinfo {author} {\bibfnamefont {O.~R.}\
  \bibnamefont {Albertini}}, \ and\ \bibinfo {author} {\bibfnamefont {A.~Y.}\
  \bibnamefont {Liu}},\ }\href {\doibase 10.1103/PhysRevB.99.174113} {\bibfield
   {journal} {\bibinfo  {journal} {Phys. Rev. B}\ }\textbf {\bibinfo {volume}
  {99}},\ \bibinfo {pages} {174113} (\bibinfo {year} {2019})}\BibitemShut
  {NoStop}%
\bibitem [{\citenamefont {Jia}\ \emph {et~al.}(2018)\citenamefont {Jia},
  \citenamefont {Rebec}, \citenamefont {Tang}, \citenamefont {Xu},
  \citenamefont {Sohail}, \citenamefont {Hashimoto}, \citenamefont {Lu},
  \citenamefont {Moore},\ and\ \citenamefont {Shen}}]{jia2018epitaxial}%
  \BibitemOpen
  \bibfield  {author} {\bibinfo {author} {\bibfnamefont {T.}~\bibnamefont
  {Jia}}, \bibinfo {author} {\bibfnamefont {S.~N.}\ \bibnamefont {Rebec}},
  \bibinfo {author} {\bibfnamefont {S.}~\bibnamefont {Tang}}, \bibinfo {author}
  {\bibfnamefont {K.}~\bibnamefont {Xu}}, \bibinfo {author} {\bibfnamefont
  {H.~M.}\ \bibnamefont {Sohail}}, \bibinfo {author} {\bibfnamefont
  {M.}~\bibnamefont {Hashimoto}}, \bibinfo {author} {\bibfnamefont
  {D.}~\bibnamefont {Lu}}, \bibinfo {author} {\bibfnamefont {R.~G.}\
  \bibnamefont {Moore}}, \ and\ \bibinfo {author} {\bibfnamefont {Z.~X.}\
  \bibnamefont {Shen}},\ }\href {\doibase 10.1088/2053-1583/aaeadf} {\bibfield
  {journal} {\bibinfo  {journal} {2D Materials}\ }\textbf {\bibinfo {volume}
  {6}},\ \bibinfo {pages} {011008} (\bibinfo {year} {2018})}\BibitemShut
  {NoStop}%
\bibitem [{\citenamefont {Rice}\ and\ \citenamefont
  {Scott}(1975)}]{Rice1975new}%
  \BibitemOpen
  \bibfield  {author} {\bibinfo {author} {\bibfnamefont {T.~M.}\ \bibnamefont
  {Rice}}\ and\ \bibinfo {author} {\bibfnamefont {G.~K.}\ \bibnamefont
  {Scott}},\ }\href {\doibase 10.1103/PhysRevLett.35.120} {\bibfield  {journal}
  {\bibinfo  {journal} {Phys. Rev. Lett.}\ }\textbf {\bibinfo {volume} {35}},\
  \bibinfo {pages} {120} (\bibinfo {year} {1975})}\BibitemShut {NoStop}%
\bibitem [{\citenamefont {Wilson}(1977)}]{wilson1977concerning}%
  \BibitemOpen
  \bibfield  {author} {\bibinfo {author} {\bibfnamefont {J.~A.}\ \bibnamefont
  {Wilson}},\ }\href {\doibase 10.1016/0038-1098(77)90133-8} {\bibfield
  {journal} {\bibinfo  {journal} {Solid State Commun.}\ }\textbf {\bibinfo
  {volume} {22}},\ \bibinfo {pages} {551} (\bibinfo {year} {1977})}\BibitemShut
  {NoStop}%
\bibitem [{\citenamefont {Kogar}\ \emph {et~al.}(2017)\citenamefont {Kogar},
  \citenamefont {Rak}, \citenamefont {Vig}, \citenamefont {Husain},
  \citenamefont {Flicker}, \citenamefont {Joe}, \citenamefont {Venema},
  \citenamefont {MacDougall}, \citenamefont {Chiang}, \citenamefont {Fradkin},
  \citenamefont {van Wezel},\ and\ \citenamefont
  {Abbamonte}}]{kogar2017signatures}%
  \BibitemOpen
  \bibfield  {author} {\bibinfo {author} {\bibfnamefont {A.}~\bibnamefont
  {Kogar}}, \bibinfo {author} {\bibfnamefont {M.~S.}\ \bibnamefont {Rak}},
  \bibinfo {author} {\bibfnamefont {S.}~\bibnamefont {Vig}}, \bibinfo {author}
  {\bibfnamefont {A.~A.}\ \bibnamefont {Husain}}, \bibinfo {author}
  {\bibfnamefont {F.}~\bibnamefont {Flicker}}, \bibinfo {author} {\bibfnamefont
  {Y.~I.}\ \bibnamefont {Joe}}, \bibinfo {author} {\bibfnamefont
  {L.}~\bibnamefont {Venema}}, \bibinfo {author} {\bibfnamefont {G.~J.}\
  \bibnamefont {MacDougall}}, \bibinfo {author} {\bibfnamefont {T.~C.}\
  \bibnamefont {Chiang}}, \bibinfo {author} {\bibfnamefont {E.}~\bibnamefont
  {Fradkin}}, \bibinfo {author} {\bibfnamefont {J.}~\bibnamefont {van Wezel}},
  \ and\ \bibinfo {author} {\bibfnamefont {P.}~\bibnamefont {Abbamonte}},\
  }\href {\doibase 10.1126/science.aam6432} {\bibfield  {journal} {\bibinfo
  {journal} {Science}\ }\textbf {\bibinfo {volume} {358}},\ \bibinfo {pages}
  {1314} (\bibinfo {year} {2017})}\BibitemShut {NoStop}%
\bibitem [{\citenamefont {Whangbo}\ and\ \citenamefont
  {Canadell}(1992)}]{whangbo1992analogies}%
  \BibitemOpen
  \bibfield  {author} {\bibinfo {author} {\bibfnamefont {M.~H.}\ \bibnamefont
  {Whangbo}}\ and\ \bibinfo {author} {\bibfnamefont {E.}~\bibnamefont
  {Canadell}},\ }\href {\doibase 10.1021/ja00050a044} {\bibfield  {journal}
  {\bibinfo  {journal} {J. Am. Chem. Soc.}\ }\textbf {\bibinfo {volume}
  {114}},\ \bibinfo {pages} {9587} (\bibinfo {year} {1992})}\BibitemShut
  {NoStop}%
\bibitem [{\citenamefont {Kidd}\ \emph {et~al.}(2002)\citenamefont {Kidd},
  \citenamefont {Miller}, \citenamefont {Chou},\ and\ \citenamefont
  {Chiang}}]{Kidd2002electron}%
  \BibitemOpen
  \bibfield  {author} {\bibinfo {author} {\bibfnamefont {T.~E.}\ \bibnamefont
  {Kidd}}, \bibinfo {author} {\bibfnamefont {T.}~\bibnamefont {Miller}},
  \bibinfo {author} {\bibfnamefont {M.~Y.}\ \bibnamefont {Chou}}, \ and\
  \bibinfo {author} {\bibfnamefont {T.~C.}\ \bibnamefont {Chiang}},\ }\href
  {\doibase 10.1103/PhysRevLett.88.226402} {\bibfield  {journal} {\bibinfo
  {journal} {Phys. Rev. Lett.}\ }\textbf {\bibinfo {volume} {88}},\ \bibinfo
  {pages} {226402} (\bibinfo {year} {2002})}\BibitemShut {NoStop}%
\bibitem [{\citenamefont {Guo}\ \emph {et~al.}(2016)\citenamefont {Guo},
  \citenamefont {Tian}, \citenamefont {Xiao}, \citenamefont {Mi},\ and\
  \citenamefont {Xue}}]{guo2016field}%
  \BibitemOpen
  \bibfield  {author} {\bibinfo {author} {\bibfnamefont {C.}~\bibnamefont
  {Guo}}, \bibinfo {author} {\bibfnamefont {Z.}~\bibnamefont {Tian}}, \bibinfo
  {author} {\bibfnamefont {Y.}~\bibnamefont {Xiao}}, \bibinfo {author}
  {\bibfnamefont {Q.}~\bibnamefont {Mi}}, \ and\ \bibinfo {author}
  {\bibfnamefont {J.}~\bibnamefont {Xue}},\ }\href {\doibase 10.1063/1.4967744}
  {\bibfield  {journal} {\bibinfo  {journal} {Appl. Phys. Lett.}\ }\textbf
  {\bibinfo {volume} {109}},\ \bibinfo {pages} {203104} (\bibinfo {year}
  {2016})}\BibitemShut {NoStop}%
\bibitem [{\citenamefont {Zhou}\ \emph {et~al.}(2015)\citenamefont {Zhou},
  \citenamefont {Gan}, \citenamefont {Tian}, \citenamefont {Zhang},
  \citenamefont {Jin}, \citenamefont {Li}, \citenamefont {Bando}, \citenamefont
  {Golberg},\ and\ \citenamefont {Zhai}}]{zhou2015ultrathin}%
  \BibitemOpen
  \bibfield  {author} {\bibinfo {author} {\bibfnamefont {X.}~\bibnamefont
  {Zhou}}, \bibinfo {author} {\bibfnamefont {L.}~\bibnamefont {Gan}}, \bibinfo
  {author} {\bibfnamefont {W.}~\bibnamefont {Tian}}, \bibinfo {author}
  {\bibfnamefont {Q.}~\bibnamefont {Zhang}}, \bibinfo {author} {\bibfnamefont
  {S.}~\bibnamefont {Jin}}, \bibinfo {author} {\bibfnamefont {H.}~\bibnamefont
  {Li}}, \bibinfo {author} {\bibfnamefont {Y.}~\bibnamefont {Bando}}, \bibinfo
  {author} {\bibfnamefont {D.}~\bibnamefont {Golberg}}, \ and\ \bibinfo
  {author} {\bibfnamefont {T.}~\bibnamefont {Zhai}},\ }\href {\doibase
  10.1002/adma.201503873} {\bibfield  {journal} {\bibinfo  {journal} {Adv.
  Mater.}\ }\textbf {\bibinfo {volume} {27}},\ \bibinfo {pages} {8035}
  (\bibinfo {year} {2015})}\BibitemShut {NoStop}%
\bibitem [{\citenamefont {Huang}\ \emph {et~al.}(2015)\citenamefont {Huang},
  \citenamefont {Xu}, \citenamefont {Wang}, \citenamefont {Shifa},
  \citenamefont {Wang}, \citenamefont {Wang}, \citenamefont {Jiang},\ and\
  \citenamefont {He}}]{huang2015designing}%
  \BibitemOpen
  \bibfield  {author} {\bibinfo {author} {\bibfnamefont {Y.}~\bibnamefont
  {Huang}}, \bibinfo {author} {\bibfnamefont {K.}~\bibnamefont {Xu}}, \bibinfo
  {author} {\bibfnamefont {Z.}~\bibnamefont {Wang}}, \bibinfo {author}
  {\bibfnamefont {T.~A.}\ \bibnamefont {Shifa}}, \bibinfo {author}
  {\bibfnamefont {Q.}~\bibnamefont {Wang}}, \bibinfo {author} {\bibfnamefont
  {F.}~\bibnamefont {Wang}}, \bibinfo {author} {\bibfnamefont {C.}~\bibnamefont
  {Jiang}}, \ and\ \bibinfo {author} {\bibfnamefont {J.}~\bibnamefont {He}},\
  }\href {\doibase 10.1039/C5NR05989E} {\bibfield  {journal} {\bibinfo
  {journal} {Nanoscale}\ }\textbf {\bibinfo {volume} {7}},\ \bibinfo {pages}
  {17375} (\bibinfo {year} {2015})}\BibitemShut {NoStop}%
\bibitem [{\citenamefont {Luo}\ \emph {et~al.}(2018)\citenamefont {Luo},
  \citenamefont {Zheng}, \citenamefont {Luo}, \citenamefont {Hao},
  \citenamefont {Du}, \citenamefont {Liang}, \citenamefont {Li}, \citenamefont
  {Khor}, \citenamefont {Hippalgaonkar}, \citenamefont {Xu}, \citenamefont
  {Yan}, \citenamefont {Wolverton},\ and\ \citenamefont
  {Kanatzidis}}]{luo2018n}%
  \BibitemOpen
  \bibfield  {author} {\bibinfo {author} {\bibfnamefont {Y.~B.}\ \bibnamefont
  {Luo}}, \bibinfo {author} {\bibfnamefont {Y.}~\bibnamefont {Zheng}}, \bibinfo
  {author} {\bibfnamefont {Z.~Z.}\ \bibnamefont {Luo}}, \bibinfo {author}
  {\bibfnamefont {S.~Q.}\ \bibnamefont {Hao}}, \bibinfo {author} {\bibfnamefont
  {C.~F.}\ \bibnamefont {Du}}, \bibinfo {author} {\bibfnamefont {Q.~H.}\
  \bibnamefont {Liang}}, \bibinfo {author} {\bibfnamefont {Z.}~\bibnamefont
  {Li}}, \bibinfo {author} {\bibfnamefont {K.~A.}\ \bibnamefont {Khor}},
  \bibinfo {author} {\bibfnamefont {K.}~\bibnamefont {Hippalgaonkar}}, \bibinfo
  {author} {\bibfnamefont {J.~W.}\ \bibnamefont {Xu}}, \bibinfo {author}
  {\bibfnamefont {Q.~Y.}\ \bibnamefont {Yan}}, \bibinfo {author} {\bibfnamefont
  {C.}~\bibnamefont {Wolverton}}, \ and\ \bibinfo {author} {\bibfnamefont
  {M.~G.}\ \bibnamefont {Kanatzidis}},\ }\href {\doibase
  10.1002/aenm.201702167} {\bibfield  {journal} {\bibinfo  {journal} {Adv.
  Energy Mater.}\ }\textbf {\bibinfo {volume} {8}},\ \bibinfo {pages} {1702167}
  (\bibinfo {year} {2018})}\BibitemShut {NoStop}%
\bibitem [{\citenamefont {Wu}\ \emph {et~al.}(2019)\citenamefont {Wu},
  \citenamefont {Li}, \citenamefont {Susner}, \citenamefont {Kwon},
  \citenamefont {Kim}, \citenamefont {Haugan},\ and\ \citenamefont
  {Lv}}]{wu2019spacing}%
  \BibitemOpen
  \bibfield  {author} {\bibinfo {author} {\bibfnamefont {H.}~\bibnamefont
  {Wu}}, \bibinfo {author} {\bibfnamefont {S.}~\bibnamefont {Li}}, \bibinfo
  {author} {\bibfnamefont {M.}~\bibnamefont {Susner}}, \bibinfo {author}
  {\bibfnamefont {S.}~\bibnamefont {Kwon}}, \bibinfo {author} {\bibfnamefont
  {M.}~\bibnamefont {Kim}}, \bibinfo {author} {\bibfnamefont {T.}~\bibnamefont
  {Haugan}}, \ and\ \bibinfo {author} {\bibfnamefont {B.}~\bibnamefont {Lv}},\
  }\href {\doibase 10.1088/2053-1583/ab3ea1} {\bibfield  {journal} {\bibinfo
  {journal} {2D Mater.}\ }\textbf {\bibinfo {volume} {6}},\ \bibinfo {pages}
  {045048} (\bibinfo {year} {2019})}\BibitemShut {NoStop}%
\bibitem [{\citenamefont {Song}\ \emph {et~al.}(2019)\citenamefont {Song},
  \citenamefont {Liang}, \citenamefont {Guo}, \citenamefont {Deng},
  \citenamefont {Gao},\ and\ \citenamefont {Chen}}]{Song2019superconductivity}%
  \BibitemOpen
  \bibfield  {author} {\bibinfo {author} {\bibfnamefont {Y.~P.}\ \bibnamefont
  {Song}}, \bibinfo {author} {\bibfnamefont {X.~W.}\ \bibnamefont {Liang}},
  \bibinfo {author} {\bibfnamefont {J.~G.}\ \bibnamefont {Guo}}, \bibinfo
  {author} {\bibfnamefont {J.}~\bibnamefont {Deng}}, \bibinfo {author}
  {\bibfnamefont {G.~Y.}\ \bibnamefont {Gao}}, \ and\ \bibinfo {author}
  {\bibfnamefont {X.~L.}\ \bibnamefont {Chen}},\ }\href {\doibase
  10.1103/PhysRevMaterials.3.054804} {\bibfield  {journal} {\bibinfo  {journal}
  {Phys. Rev. Materials}\ }\textbf {\bibinfo {volume} {3}},\ \bibinfo {pages}
  {054804} (\bibinfo {year} {2019})}\BibitemShut {NoStop}%
\bibitem [{\citenamefont {Zhang}\ \emph
  {et~al.}(2018{\natexlab{a}})\citenamefont {Zhang}, \citenamefont {Fan},
  \citenamefont {Wang}, \citenamefont {Zhang}, \citenamefont {Wang},
  \citenamefont {Li}, \citenamefont {He}, \citenamefont {Song}, \citenamefont
  {Ma},\ and\ \citenamefont {Xue}}]{Zhang2018observation}%
  \BibitemOpen
  \bibfield  {author} {\bibinfo {author} {\bibfnamefont {Y.~M.}\ \bibnamefont
  {Zhang}}, \bibinfo {author} {\bibfnamefont {J.~Q.}\ \bibnamefont {Fan}},
  \bibinfo {author} {\bibfnamefont {W.~L.}\ \bibnamefont {Wang}}, \bibinfo
  {author} {\bibfnamefont {D.}~\bibnamefont {Zhang}}, \bibinfo {author}
  {\bibfnamefont {L.}~\bibnamefont {Wang}}, \bibinfo {author} {\bibfnamefont
  {W.}~\bibnamefont {Li}}, \bibinfo {author} {\bibfnamefont {K.}~\bibnamefont
  {He}}, \bibinfo {author} {\bibfnamefont {C.~L.}\ \bibnamefont {Song}},
  \bibinfo {author} {\bibfnamefont {X.~C.}\ \bibnamefont {Ma}}, \ and\ \bibinfo
  {author} {\bibfnamefont {Q.~K.}\ \bibnamefont {Xue}},\ }\href {\doibase
  10.1103/PhysRevB.98.220508} {\bibfield  {journal} {\bibinfo  {journal} {Phys.
  Rev. B}\ }\textbf {\bibinfo {volume} {98}},\ \bibinfo {pages} {220508}
  (\bibinfo {year} {2018}{\natexlab{a}})}\BibitemShut {NoStop}%
\bibitem [{\citenamefont {Shao}\ \emph {et~al.}(2019)\citenamefont {Shao},
  \citenamefont {Fu}, \citenamefont {Li}, \citenamefont {Cao}, \citenamefont
  {Bian}, \citenamefont {Sun}, \citenamefont {Zhang}, \citenamefont {Gedeon},
  \citenamefont {Zhang}, \citenamefont {Liu}, \citenamefont {Cheng},
  \citenamefont {Zheng}, \citenamefont {Zhang},\ and\ \citenamefont
  {Pan}}]{shao2019strongly}%
  \BibitemOpen
  \bibfield  {author} {\bibinfo {author} {\bibfnamefont {Z.~B.}\ \bibnamefont
  {Shao}}, \bibinfo {author} {\bibfnamefont {Z.~G.}\ \bibnamefont {Fu}},
  \bibinfo {author} {\bibfnamefont {S.~J.}\ \bibnamefont {Li}}, \bibinfo
  {author} {\bibfnamefont {Y.}~\bibnamefont {Cao}}, \bibinfo {author}
  {\bibfnamefont {Q.}~\bibnamefont {Bian}}, \bibinfo {author} {\bibfnamefont
  {H.}~\bibnamefont {Sun}}, \bibinfo {author} {\bibfnamefont {Z.~Y.}\
  \bibnamefont {Zhang}}, \bibinfo {author} {\bibfnamefont {H.}~\bibnamefont
  {Gedeon}}, \bibinfo {author} {\bibfnamefont {X.}~\bibnamefont {Zhang}},
  \bibinfo {author} {\bibfnamefont {L.~L.~J.}\ \bibnamefont {Liu}}, \bibinfo
  {author} {\bibfnamefont {Z.~W.}\ \bibnamefont {Cheng}}, \bibinfo {author}
  {\bibfnamefont {F.~W.}\ \bibnamefont {Zheng}}, \bibinfo {author}
  {\bibfnamefont {P.}~\bibnamefont {Zhang}}, \ and\ \bibinfo {author}
  {\bibfnamefont {M.~H.}\ \bibnamefont {Pan}},\ }\href {\doibase
  10.1021/acs.nanolett.9b01766} {\bibfield  {journal} {\bibinfo  {journal}
  {Nano Lett.}\ }\textbf {\bibinfo {volume} {19}},\ \bibinfo {pages} {5304}
  (\bibinfo {year} {2019})}\BibitemShut {NoStop}%
\bibitem [{\citenamefont {Zeng}\ \emph {et~al.}(2018)\citenamefont {Zeng},
  \citenamefont {Liu}, \citenamefont {Fu}, \citenamefont {Chen}, \citenamefont
  {Pan}, \citenamefont {Wang}, \citenamefont {Wang}, \citenamefont {Wang},
  \citenamefont {Xu}, \citenamefont {Cai}, \citenamefont {Yan}, \citenamefont
  {Wang}, \citenamefont {Liu}, \citenamefont {Wang}, \citenamefont {Liang},
  \citenamefont {Cui}, \citenamefont {Hwang}, \citenamefont {Yuan},\ and\
  \citenamefont {Miao}}]{zeng2018gate}%
  \BibitemOpen
  \bibfield  {author} {\bibinfo {author} {\bibfnamefont {J.~W.}\ \bibnamefont
  {Zeng}}, \bibinfo {author} {\bibfnamefont {E.~F.}\ \bibnamefont {Liu}},
  \bibinfo {author} {\bibfnamefont {Y.~J.}\ \bibnamefont {Fu}}, \bibinfo
  {author} {\bibfnamefont {Z.~Y.}\ \bibnamefont {Chen}}, \bibinfo {author}
  {\bibfnamefont {C.}~\bibnamefont {Pan}}, \bibinfo {author} {\bibfnamefont
  {C.~Y.}\ \bibnamefont {Wang}}, \bibinfo {author} {\bibfnamefont
  {M.}~\bibnamefont {Wang}}, \bibinfo {author} {\bibfnamefont {Y.~J.}\
  \bibnamefont {Wang}}, \bibinfo {author} {\bibfnamefont {K.}~\bibnamefont
  {Xu}}, \bibinfo {author} {\bibfnamefont {S.~H.}\ \bibnamefont {Cai}},
  \bibinfo {author} {\bibfnamefont {X.~X.}\ \bibnamefont {Yan}}, \bibinfo
  {author} {\bibfnamefont {Y.}~\bibnamefont {Wang}}, \bibinfo {author}
  {\bibfnamefont {X.~W.}\ \bibnamefont {Liu}}, \bibinfo {author} {\bibfnamefont
  {P.}~\bibnamefont {Wang}}, \bibinfo {author} {\bibfnamefont {S.~J.}\
  \bibnamefont {Liang}}, \bibinfo {author} {\bibfnamefont {Y.}~\bibnamefont
  {Cui}}, \bibinfo {author} {\bibfnamefont {H.~Y.}\ \bibnamefont {Hwang}},
  \bibinfo {author} {\bibfnamefont {H.~T.}\ \bibnamefont {Yuan}}, \ and\
  \bibinfo {author} {\bibfnamefont {F.}~\bibnamefont {Miao}},\ }\href {\doibase
  10.1021/acs.nanolett.7b05157} {\bibfield  {journal} {\bibinfo  {journal}
  {Nano Lett.}\ }\textbf {\bibinfo {volume} {18}},\ \bibinfo {pages} {1410}
  (\bibinfo {year} {2018})}\BibitemShut {NoStop}%
\bibitem [{\citenamefont {Gonzalez}\ and\ \citenamefont
  {Oleynik}(2016)}]{Gonzalez2016layer}%
  \BibitemOpen
  \bibfield  {author} {\bibinfo {author} {\bibfnamefont {J.~M.}\ \bibnamefont
  {Gonzalez}}\ and\ \bibinfo {author} {\bibfnamefont {I.~I.}\ \bibnamefont
  {Oleynik}},\ }\href {\doibase 10.1103/PhysRevB.94.125443} {\bibfield
  {journal} {\bibinfo  {journal} {Phys. Rev. B}\ }\textbf {\bibinfo {volume}
  {94}},\ \bibinfo {pages} {125443} (\bibinfo {year} {2016})}\BibitemShut
  {NoStop}%
\bibitem [{\citenamefont {Ying}\ \emph {et~al.}(2018)\citenamefont {Ying},
  \citenamefont {Paudyal}, \citenamefont {Heil}, \citenamefont {Chen},
  \citenamefont {Struzhkin},\ and\ \citenamefont {Margine}}]{Ying2018unusual}%
  \BibitemOpen
  \bibfield  {author} {\bibinfo {author} {\bibfnamefont {J.~J.}\ \bibnamefont
  {Ying}}, \bibinfo {author} {\bibfnamefont {H.}~\bibnamefont {Paudyal}},
  \bibinfo {author} {\bibfnamefont {C.}~\bibnamefont {Heil}}, \bibinfo {author}
  {\bibfnamefont {X.-J.}\ \bibnamefont {Chen}}, \bibinfo {author}
  {\bibfnamefont {V.~V.}\ \bibnamefont {Struzhkin}}, \ and\ \bibinfo {author}
  {\bibfnamefont {E.~R.}\ \bibnamefont {Margine}},\ }\href {\doibase
  10.1103/PhysRevLett.121.027003} {\bibfield  {journal} {\bibinfo  {journal}
  {Phys. Rev. Lett.}\ }\textbf {\bibinfo {volume} {121}},\ \bibinfo {pages}
  {027003} (\bibinfo {year} {2018})}\BibitemShut {NoStop}%
\bibitem [{\citenamefont {Song}\ \emph {et~al.}(2011)\citenamefont {Song},
  \citenamefont {Wang}, \citenamefont {Jiang}, \citenamefont {Li},
  \citenamefont {Wang}, \citenamefont {He}, \citenamefont {Chen}, \citenamefont
  {Ma},\ and\ \citenamefont {Xue}}]{song2011molecular}%
  \BibitemOpen
  \bibfield  {author} {\bibinfo {author} {\bibfnamefont {C.~L.}\ \bibnamefont
  {Song}}, \bibinfo {author} {\bibfnamefont {Y.~L.}\ \bibnamefont {Wang}},
  \bibinfo {author} {\bibfnamefont {Y.~P.}\ \bibnamefont {Jiang}}, \bibinfo
  {author} {\bibfnamefont {Z.}~\bibnamefont {Li}}, \bibinfo {author}
  {\bibfnamefont {L.}~\bibnamefont {Wang}}, \bibinfo {author} {\bibfnamefont
  {K.}~\bibnamefont {He}}, \bibinfo {author} {\bibfnamefont {X.}~\bibnamefont
  {Chen}}, \bibinfo {author} {\bibfnamefont {X.~C.}\ \bibnamefont {Ma}}, \ and\
  \bibinfo {author} {\bibfnamefont {Q.~K.}\ \bibnamefont {Xue}},\ }\href
  {\doibase 10.1103/PhysRevB.84.020503} {\bibfield  {journal} {\bibinfo
  {journal} {Phys. Rev. B}\ }\textbf {\bibinfo {volume} {84}},\ \bibinfo
  {pages} {020503} (\bibinfo {year} {2011})}\BibitemShut {NoStop}%
\bibitem [{sup()}]{supplementary}%
  \BibitemOpen
  \href@noop {} {\bibinfo  {journal} {See Supplemental Material at for details
  regarding substrates, FFT images and CDW wave vectors.}\ }\BibitemShut
  {NoStop}%
\bibitem [{\citenamefont {Modesti}\ \emph {et~al.}(2007)\citenamefont
  {Modesti}, \citenamefont {Petaccia}, \citenamefont {Ceballos}, \citenamefont
  {Vobornik}, \citenamefont {Panaccione}, \citenamefont {Rossi}, \citenamefont
  {Ottaviano}, \citenamefont {Larciprete}, \citenamefont {Lizzit},\ and\
  \citenamefont {Goldoni}}]{Modesti2007insulating}%
  \BibitemOpen
\bibfield  {journal} {  }\bibfield  {author} {\bibinfo {author} {\bibfnamefont
  {S.}~\bibnamefont {Modesti}}, \bibinfo {author} {\bibfnamefont
  {L.}~\bibnamefont {Petaccia}}, \bibinfo {author} {\bibfnamefont
  {G.}~\bibnamefont {Ceballos}}, \bibinfo {author} {\bibfnamefont
  {I.}~\bibnamefont {Vobornik}}, \bibinfo {author} {\bibfnamefont
  {G.}~\bibnamefont {Panaccione}}, \bibinfo {author} {\bibfnamefont
  {G.}~\bibnamefont {Rossi}}, \bibinfo {author} {\bibfnamefont
  {L.}~\bibnamefont {Ottaviano}}, \bibinfo {author} {\bibfnamefont
  {R.}~\bibnamefont {Larciprete}}, \bibinfo {author} {\bibfnamefont
  {S.}~\bibnamefont {Lizzit}}, \ and\ \bibinfo {author} {\bibfnamefont
  {A.}~\bibnamefont {Goldoni}},\ }\href {\doibase
  10.1103/PhysRevLett.98.126401} {\bibfield  {journal} {\bibinfo  {journal}
  {Phys. Rev. Lett.}\ }\textbf {\bibinfo {volume} {98}},\ \bibinfo {pages}
  {126401} (\bibinfo {year} {2007})}\BibitemShut {NoStop}%
\bibitem [{\citenamefont {Wu}\ \emph {et~al.}(2020)\citenamefont {Wu},
  \citenamefont {Ming}, \citenamefont {Smith}, \citenamefont {Liu},
  \citenamefont {Ye}, \citenamefont {Wang}, \citenamefont {Johnston},\ and\
  \citenamefont {Weitering}}]{Wu2020superconductivity}%
  \BibitemOpen
  \bibfield  {author} {\bibinfo {author} {\bibfnamefont {X.~F.}\ \bibnamefont
  {Wu}}, \bibinfo {author} {\bibfnamefont {F.~F.}\ \bibnamefont {Ming}},
  \bibinfo {author} {\bibfnamefont {T.~S.}\ \bibnamefont {Smith}}, \bibinfo
  {author} {\bibfnamefont {G.}~\bibnamefont {Liu}}, \bibinfo {author}
  {\bibfnamefont {F.}~\bibnamefont {Ye}}, \bibinfo {author} {\bibfnamefont
  {K.}~\bibnamefont {Wang}}, \bibinfo {author} {\bibfnamefont {S.}~\bibnamefont
  {Johnston}}, \ and\ \bibinfo {author} {\bibfnamefont {H.~H.}\ \bibnamefont
  {Weitering}},\ }\href {\doibase 10.1103/PhysRevLett.125.117001} {\bibfield
  {journal} {\bibinfo  {journal} {Phys. Rev. Lett.}\ }\textbf {\bibinfo
  {volume} {125}},\ \bibinfo {pages} {117001} (\bibinfo {year}
  {2020})}\BibitemShut {NoStop}%
\bibitem [{\citenamefont {Lochocki}\ \emph {et~al.}(2019)\citenamefont
  {Lochocki}, \citenamefont {Vishwanath}, \citenamefont {Liu}, \citenamefont
  {Dobrowolska}, \citenamefont {Furdyna}, \citenamefont {Xing},\ and\
  \citenamefont {Shen}}]{lochocki2019electronic}%
  \BibitemOpen
  \bibfield  {author} {\bibinfo {author} {\bibfnamefont {E.~B.}\ \bibnamefont
  {Lochocki}}, \bibinfo {author} {\bibfnamefont {S.}~\bibnamefont
  {Vishwanath}}, \bibinfo {author} {\bibfnamefont {X.~Y.}\ \bibnamefont {Liu}},
  \bibinfo {author} {\bibfnamefont {M.}~\bibnamefont {Dobrowolska}}, \bibinfo
  {author} {\bibfnamefont {J.}~\bibnamefont {Furdyna}}, \bibinfo {author}
  {\bibfnamefont {H.~G.}\ \bibnamefont {Xing}}, \ and\ \bibinfo {author}
  {\bibfnamefont {K.~M.}\ \bibnamefont {Shen}},\ }\href {\doibase
  10.1063/1.5084147} {\bibfield  {journal} {\bibinfo  {journal} {Appl. Phys.
  Lett.}\ }\textbf {\bibinfo {volume} {114}},\ \bibinfo {pages} {091602}
  (\bibinfo {year} {2019})}\BibitemShut {NoStop}%
\bibitem [{\citenamefont {Zhang}\ \emph
  {et~al.}(2018{\natexlab{b}})\citenamefont {Zhang}, \citenamefont {Li},
  \citenamefont {Lochocki}, \citenamefont {Vishwanath}, \citenamefont {Liu},
  \citenamefont {Yan}, \citenamefont {Lien}, \citenamefont {Dobrowolska},
  \citenamefont {Furdyna}, \citenamefont {Shen}, \citenamefont {Cheng},
  \citenamefont {Hight~Walker}, \citenamefont {Gundlach}, \citenamefont
  {Xing},\ and\ \citenamefont {Nghyen}}]{zhang2018band}%
  \BibitemOpen
  \bibfield  {author} {\bibinfo {author} {\bibfnamefont {Q.}~\bibnamefont
  {Zhang}}, \bibinfo {author} {\bibfnamefont {M.}~\bibnamefont {Li}}, \bibinfo
  {author} {\bibfnamefont {E.~B.}\ \bibnamefont {Lochocki}}, \bibinfo {author}
  {\bibfnamefont {S.}~\bibnamefont {Vishwanath}}, \bibinfo {author}
  {\bibfnamefont {X.}~\bibnamefont {Liu}}, \bibinfo {author} {\bibfnamefont
  {R.}~\bibnamefont {Yan}}, \bibinfo {author} {\bibfnamefont {H.-H.}\
  \bibnamefont {Lien}}, \bibinfo {author} {\bibfnamefont {M.}~\bibnamefont
  {Dobrowolska}}, \bibinfo {author} {\bibfnamefont {J.}~\bibnamefont
  {Furdyna}}, \bibinfo {author} {\bibfnamefont {K.~M.}\ \bibnamefont {Shen}},
  \bibinfo {author} {\bibfnamefont {G.~J.}\ \bibnamefont {Cheng}}, \bibinfo
  {author} {\bibfnamefont {A.~R.}\ \bibnamefont {Hight~Walker}}, \bibinfo
  {author} {\bibfnamefont {D.~J.}\ \bibnamefont {Gundlach}}, \bibinfo {author}
  {\bibfnamefont {H.~G.}\ \bibnamefont {Xing}}, \ and\ \bibinfo {author}
  {\bibfnamefont {N.~V.}\ \bibnamefont {Nghyen}},\ }\href {\doibase
  10.1063/1.5016183} {\bibfield  {journal} {\bibinfo  {journal} {Appl. Phys.
  Lett.}\ }\textbf {\bibinfo {volume} {112}},\ \bibinfo {pages} {042108}
  (\bibinfo {year} {2018}{\natexlab{b}})}\BibitemShut {NoStop}%
\bibitem [{\citenamefont {L{\"u}th}(2001)}]{luth2001solid}%
  \BibitemOpen
  \bibfield  {author} {\bibinfo {author} {\bibfnamefont {H.}~\bibnamefont
  {L{\"u}th}},\ }\href@noop {} {\emph {\bibinfo {title} {Solid surfaces,
  interfaces and thin films}}},\ Vol.~\bibinfo {volume} {4}\ (\bibinfo
  {publisher} {Springer},\ \bibinfo {year} {2001})\BibitemShut {NoStop}%
\bibitem [{\citenamefont {Monch}(1990)}]{monch1990physics}%
  \BibitemOpen
  \bibfield  {author} {\bibinfo {author} {\bibfnamefont {W.}~\bibnamefont
  {Monch}},\ }\href {\doibase 10.1088/0034-4885/53/3/001} {\bibfield  {journal}
  {\bibinfo  {journal} {Rep. Prog. Phys.}\ }\textbf {\bibinfo {volume} {53}},\
  \bibinfo {pages} {221} (\bibinfo {year} {1990})}\BibitemShut {NoStop}%
\bibitem [{\citenamefont {Kerelsky}\ \emph {et~al.}(2017)\citenamefont
  {Kerelsky}, \citenamefont {Nipane}, \citenamefont {Edelberg}, \citenamefont
  {Wang}, \citenamefont {Zhou}, \citenamefont {Motmaendadgar}, \citenamefont
  {Gao}, \citenamefont {Xie}, \citenamefont {Kang}, \citenamefont {Park},
  \citenamefont {Teherani},\ and\ \citenamefont
  {Pasupathy}}]{kerelsky2017absence}%
  \BibitemOpen
  \bibfield  {author} {\bibinfo {author} {\bibfnamefont {A.}~\bibnamefont
  {Kerelsky}}, \bibinfo {author} {\bibfnamefont {A.}~\bibnamefont {Nipane}},
  \bibinfo {author} {\bibfnamefont {D.}~\bibnamefont {Edelberg}}, \bibinfo
  {author} {\bibfnamefont {D.}~\bibnamefont {Wang}}, \bibinfo {author}
  {\bibfnamefont {X.}~\bibnamefont {Zhou}}, \bibinfo {author} {\bibfnamefont
  {A.}~\bibnamefont {Motmaendadgar}}, \bibinfo {author} {\bibfnamefont
  {H.}~\bibnamefont {Gao}}, \bibinfo {author} {\bibfnamefont {S.}~\bibnamefont
  {Xie}}, \bibinfo {author} {\bibfnamefont {K.}~\bibnamefont {Kang}}, \bibinfo
  {author} {\bibfnamefont {J.}~\bibnamefont {Park}}, \bibinfo {author}
  {\bibfnamefont {J.}~\bibnamefont {Teherani}}, \ and\ \bibinfo {author}
  {\bibfnamefont {A.}~\bibnamefont {Pasupathy}},\ }\href {\doibase
  10.1021/acs.nanolett.7b01986} {\bibfield  {journal} {\bibinfo  {journal}
  {Nano Lett.}\ }\textbf {\bibinfo {volume} {17}},\ \bibinfo {pages} {5962}
  (\bibinfo {year} {2017})}\BibitemShut {NoStop}%
\bibitem [{\citenamefont {Rossnagel}\ \emph {et~al.}(2002)\citenamefont
  {Rossnagel}, \citenamefont {Kipp},\ and\ \citenamefont
  {Skibowski}}]{Rossnagel2002charge}%
  \BibitemOpen
  \bibfield  {author} {\bibinfo {author} {\bibfnamefont {K.}~\bibnamefont
  {Rossnagel}}, \bibinfo {author} {\bibfnamefont {L.}~\bibnamefont {Kipp}}, \
  and\ \bibinfo {author} {\bibfnamefont {M.}~\bibnamefont {Skibowski}},\ }\href
  {\doibase 10.1103/PhysRevB.65.235101} {\bibfield  {journal} {\bibinfo
  {journal} {Phys. Rev. B}\ }\textbf {\bibinfo {volume} {65}},\ \bibinfo
  {pages} {235101} (\bibinfo {year} {2002})}\BibitemShut {NoStop}%
\bibitem [{\citenamefont {Rossnagel}(2010)}]{rossnagel2010suppression}%
  \BibitemOpen
  \bibfield  {author} {\bibinfo {author} {\bibfnamefont {K.}~\bibnamefont
  {Rossnagel}},\ }\href {\doibase 10.1088/1367-2630/12/12/125018} {\bibfield
  {journal} {\bibinfo  {journal} {New J. Phys.}\ }\textbf {\bibinfo {volume}
  {12}},\ \bibinfo {pages} {125018} (\bibinfo {year} {2010})}\BibitemShut
  {NoStop}%
\bibitem [{\citenamefont {Hughes}(1977)}]{hughes1977structural}%
  \BibitemOpen
  \bibfield  {author} {\bibinfo {author} {\bibfnamefont {H.~P.}\ \bibnamefont
  {Hughes}},\ }\href {\doibase 10.1088/0022-3719/10/11/009} {\bibfield
  {journal} {\bibinfo  {journal} {J. Phys. C: Solid State Phys.}\ }\textbf
  {\bibinfo {volume} {10}},\ \bibinfo {pages} {L319} (\bibinfo {year}
  {1977})}\BibitemShut {NoStop}%
\end{thebibliography}
%

\end{document}